\documentclass[aip,jcp,preprint]{revtex4-2}
\usepackage{amssymb}
\usepackage{graphicx}% Include figure files
\usepackage[svgnames]{xcolor}
\usepackage[colorlinks=true,citecolor=blue,linkcolor=blue,citebordercolor=white,linkbordercolor=white]{hyperref}

\def\eq#1{{Eq.~(\ref{#1})}}

\begin{document}

\author{S.V. Novikov}
\email{novikov@elchem.ac.ru}
\affiliation{A.N. Frumkin Institute of
Physical Chemistry and Electrochemistry, Leninsky prosp. 31,
Moscow 119071, Russia}
\affiliation{National Research University Higher School of Economics, Myasnitskaya Ulitsa 20, Moscow 101000, Russia}

\title[DOS: emergence of the exponential tails] {Density of states in locally ordered amorphous organic semiconductors: emergence of the exponential tails}

\begin{abstract}
We present a simple model of the local order in amorphous organic semiconductors which naturally produces a spatially correlated exponential density of states (DOS). The dominant contribution to the random energy landscape is provided by electrostatic contributions from dipoles or quadrupoles. An assumption of the preferable parallel orientation of neighbor quadrupoles or antiparallel orientation of dipoles directly leads to the formation of the exponential tails of the DOS even for a moderate size of the ordered domains. The insensitivity of the exponential tail formation to the details of the microstructure of the material suggests that this mechanism is rather common in amorphous organic semiconductors.
\end{abstract}

\maketitle

\section{Introduction}
The energy density of states (DOS) is one of the most fundamental characteristics of the amorphous semiconductors. It describes the distribution density of the random carrier energy $E$ and defines or greatly influences all charge transport properties of the materials. In organic semiconductors the exact DOS $P(E)$ depends on the particular features of the material in question and typically has a complex shape. Taking into account that for the quasi-equilibrium charge transport a behavior of the low energy tail is relevant, simple DOS models emphasizing the particular shape of the tail are widely used in analytical or computer simulation analysis of the charge transport properties.

There are numerous possible sources of energetic disorder in amorphous organic semiconductors; citing just a few of them we could mention electrostatic disorder, originated from randomly located and oriented permanent dipoles and quadrupoles, \cite{Dieckmann:8136,Novikov:482,Novikov:164510} conformational disorder, \cite{Masse:115204} or mesoscopic disorder associated with various irregularities at the boundaries between microcrystals (i.e., ordered domains of organic material).\cite{Rivnay:952,Kalihari:4033,Tang:24102,Rivnay:45203,Jimison:1568} In this paper we limit our consideration to the case where the electrostatic contribution to the DOS is dominant. Detailed microscopic computer simulation suggests that this is possible in some typical organic semiconductors.\cite{Masse:115204}

A popular model of the DOS in amorphous organic materials is the Gaussian DOS.\cite{Bassler:15} Not surprisingly, the Gaussian DOS quite naturally emerges in the models where the electrostatic contribution is dominating because the long range nature of the unscreened electrostatic potential implies the applicability of the Central Limit Theorem.\cite{Dieckmann:8136,Novikov:482,Novikov:164510} If we consider the simplest case of the low carrier density and weak applied electric field, then for the thick transport layers having long transit time the quasi-equilibrium distribution of occupied states develops
\begin{equation}\label{GDOS-eq}
  P_{eq}(E)\propto P(E)\exp(-E/kT)\propto \exp\left[-\frac{\left(E-E_{eq}\right)^2}{2\sigma^2}\right], \hskip10pt E_{eq}=-\frac{\sigma^2}{kT},
\end{equation}
and the functional form of the DOS at $E\simeq E_{eq}$ determines properties of the quasi-equilibrium charge transport. For typical rms disorder $\sigma \simeq 0.1$ eV and room temperature $\sigma/kT \approx 4 - 5$, $P(E_{eq})/P(0) \simeq 10^{-6} - 10^{-4}$ and relevant transport states belong to the tail of the DOS. At the same time, tails of the distribution are typically affected by the subtle details of the material structure and, quite frequently, the shape of the distribution in this region notably deviates from the Gaussian one.

Most popular electrostatic models use the basic assumption of the absolute independent orientations of dipoles or quadrupoles, \cite{Dieckmann:8136,Novikov:482,Novikov:164510} yet it is evident that this assumption cannot be valid at least for neighbor molecules. Quite large and asymmetric molecules of organic semiconductors just cannot be oriented in a true random and independent way at the short distance. Obviously, short range correlation should affect the density of states in amorphous materials as well as the correlation properties of the random energy landscape and, hence, the charge transport behavior of organic materials. Such effects were indeed observed.\cite{Inoue:303,Thurzo:1108}

Additional argument for the local order comes from the different and rather unexpected side. Typically, rms electrostatic disorder $\sigma$ was estimated using the macroscopic value of static dielectric constant $\varepsilon$. This approximation is unreasonable for the short range distance comparable to the intermolecular distance. Qualitatively, more proper description could be achieved by setting $\varepsilon\approx 1$ for short distances,\cite{Novikov:275} leading to the greater $\sigma$ value. Indeed, it was demonstrated that the more accurate microscopic calculation provide approximately twice greater disorder, i.e. instead of $\sigma\simeq 0.1$ eV we should rather have $\sigma\simeq 0.2$ eV.\cite{Madigan:216402} Such value is unreasonably high and disagree with the experimentally obtained $\sigma$.\cite{Borsenberger:book,Schein:7295} Again, the plausible explanation for the agreement is the effect of the local order diminishing the total energetic rms disorder.

The tail of the DOS in amorphous organic materials is frequently described by the simple exponential form\cite{Vissenberg:12964,Dunlap:9076,Abdalla:241202,Xiao:24034}
\begin{equation}\label{exp}
  P(E) \propto \exp(E/E_e), \hskip10pt E \rightarrow -\infty,
\end{equation}
and this type of DOS leads to a drastically different transport behavior for low temperature $kT < E_e$, where transport becomes dispersive with no well-defined time-independent average carrier velocity. Several models of the development of the exponential or near-exponential tails in amorphous materials have been suggested and  some of then are hardly applicable to organic materials while others required the presence of charged dopants.\cite{Silver:177,Arkhipov:045214,Economou:6172,Bacalis:2714} Until now the exponential tails in amorphous organic semiconductors are mostly attributed to the effect of charged dopants.

In this paper we suggest a new and probably quite general model of the development of the exponential tails of the DOS in locally ordered organic glasses with dominating electrostatic disorder without presence of charged dopants. The model is directly applied to quadrupolar materials and may be applied to dipolar materials with the particular type of the local ordering.

\section{Model of the local order in quadrupole glass}

The quadrupole moment of a molecule is described by the symmetric traceless tensor
$\textbf{Q}$, $\sum\limits_i Q_{ii}=0$. Interaction energy for the charge and point quadrupole separated by the distance $\vec{r}$ is
\begin{equation}\label{eQ-U}
  E_q(\vec{r})=\frac{e\sum\limits_{i,j}Q_{ij}r_i r_j}{2\varepsilon r^5}.
\end{equation}
Quadrupole tensor may be transformed to the diagonal form where any quadrupole could be considered as a linear combination of axial and planar quadrupoles $\textbf{Q}=\textbf{Q}_a+ \textbf{Q}_p$, where $\textbf{Q}_a=\textrm{diag}(-Q_a/2,-Q_a/2,Q_a)$ and $\textbf{Q}_p=\textrm{diag}(Q_p,-Q_p,0)$. Even for absolutely random non-correlated distribution of quadrupoles the cases of pure axial and planar  quadrupoles are different: for planar quadrupoles the DOS is symmetric around $E=0$ while for the axial quadrupoles the tails are asymmetric. \cite{Novikov:164510} For planar quadrupoles the inversion $Q\rightarrow -Q$ that transforms the positive carrier energy to negative one and vice versa is equivalent to the spatial rotation of the quadrupole. All spatial configurations of the quadrupole have the equal weight, hence, the DOS is symmetric around $E=0$. For axial quadrupoles the inversion is not equivalent to any possible rotation and the DOS is not symmetric.

Naturally, we should also expect such difference for locally ordered materials without any preferred direction and consider the cases of axial quadrupole glass (QG) and planar QG separately.

%$\textbf{Q}=\textrm{diag}(Q(1-3\alpha)/2,-Q(1-\alpha/2),Q)$

In the simplest QG model quadrupoles occupy sites of the simple cubic lattice having lattice scale $a$ and all lengths are measured in the units of $a$. For the true amorphous organic material with random position of molecules $a$ is the distance between nearest neighbors and typically $a\simeq 1$ nm. All energies will be measured in the units of characteristic energy $E_0=eQ/2\varepsilon a^3$, where $Q$ is the axial or planar quadrupole moment.

Again, we consider the simplest possible model of the local order assuming that the amorphous organic material is built from the blocks of quadrupoles and in any particular block they have the identical orientations while the corresponding orientations in different blocks are independent and absolutely random.
The blocks are built by the Voronoi tesselation.\cite{moller2012lectures} We set $N_s$ randomly located seeding sites in the cubic basic sample having linear size of $L$ sites (with proper periodic boundary conditions) and then subdivide the basic sample into $N_s$ cells where sites of each cell are more close to one of the seeding sites than to all others. Average volume of the cell is $L^3/N_s$ sites and the typical linear size of a cell may be estimated as $l_a=L/N_s^{1/3}$ sites. For the majority of simulations we use $L=50$ and comparison with some particular simulations carried out for $L=100$ shows no significant difference.

\begin{figure}[tbp]
\includegraphics[width=3.375in]{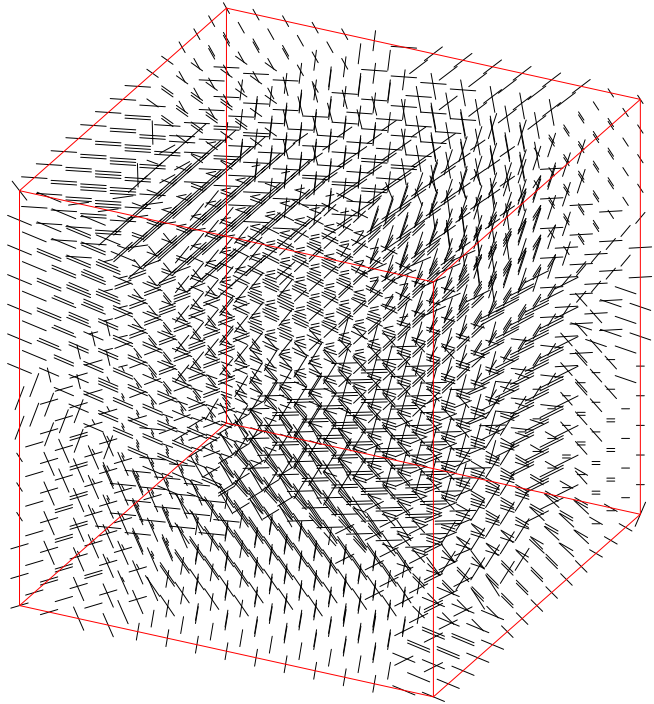}
\vskip5pt
a)

%\medskip
%\medskip
\includegraphics[width=3.375in]{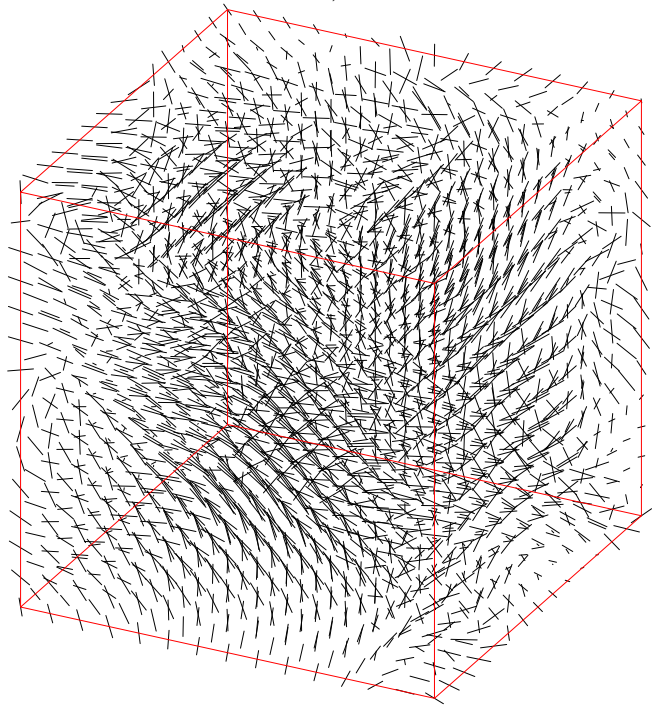}
\vskip5pt
b)
\caption{Distribution of the axes of quadrupoles for $15\times 15\times 15$ chunk of the axial Voronoi QG with $l_a=5$ for $m=0$ (a)  and $m=3$ (b). Note that due to projection effect for some sites several axes seem to belong to the single site; in fact, they are located at the different sites which are seen overlapping in projection.}  \label{axes}
\end{figure}

Boundaries of each cell are planes, thus providing sharp planar interfaces between cells. In order to check an effect of the sharp interfaces on the shape of the DOS we consider also the smoother version of the boundary by applying the following procedure: sweeping over the lattice and replacing every quadrupole with the average over nearest quadrupoles (with proper rescaling to preserve the magnitude of the quadrupole moment). In the depth of the Voronoi cell this procedure does not make any difference, while at the boundary it provides the gradual transition from one cell orientation to another (see Fig. \ref{axes}). With the increase of the number of sweeps $m$ the resulting distribution of quadrupoles orientations becomes smoother and smoother.

%fig. 1

%fig. 2
\begin{figure}[htbp]
\includegraphics[width=3.375in]{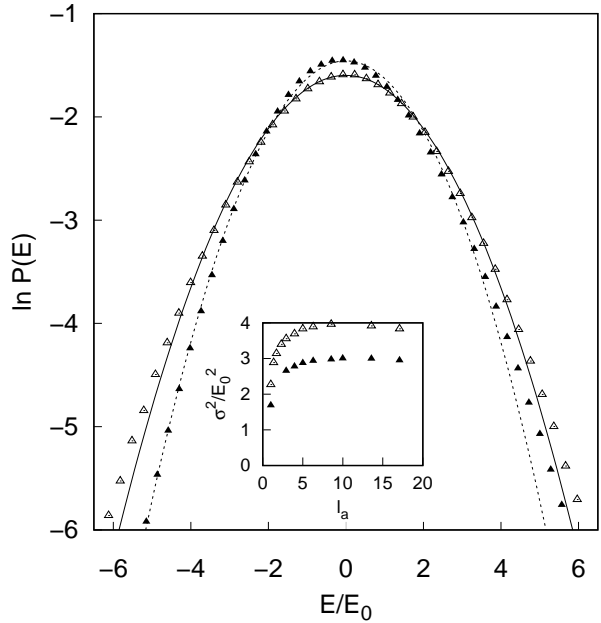}
\caption{Shape of the central region of the DOS peak for $l_a=6.3$ for the axial ($\blacktriangle$) and planar ($\vartriangle$) locally ordered QG ($L=50$). Lines show the best fit for the Gaussian function. Inset shows the dependence of $\sigma^2$ on $l_a$ for planar ($\vartriangle$) and axial ($\blacktriangle$) QG.}
\label{s2fig}
\end{figure}

%fig. 3
\begin{figure}[htbp]
\includegraphics[width=3.375in]{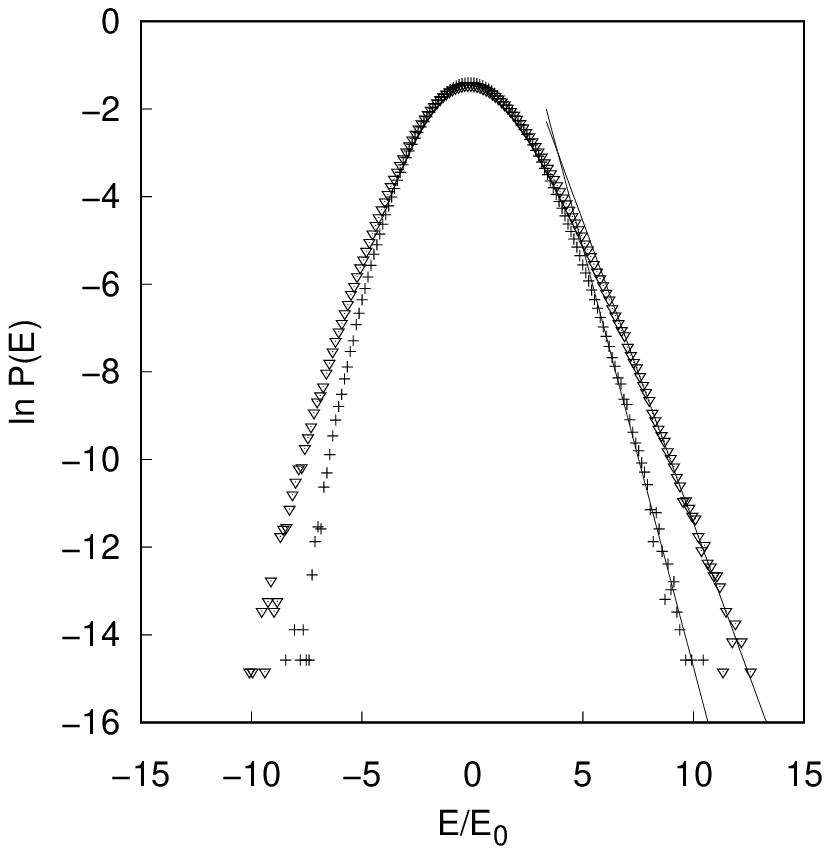}
%\vskip10pt

a)

%\medskip
\includegraphics[width=3.375in]{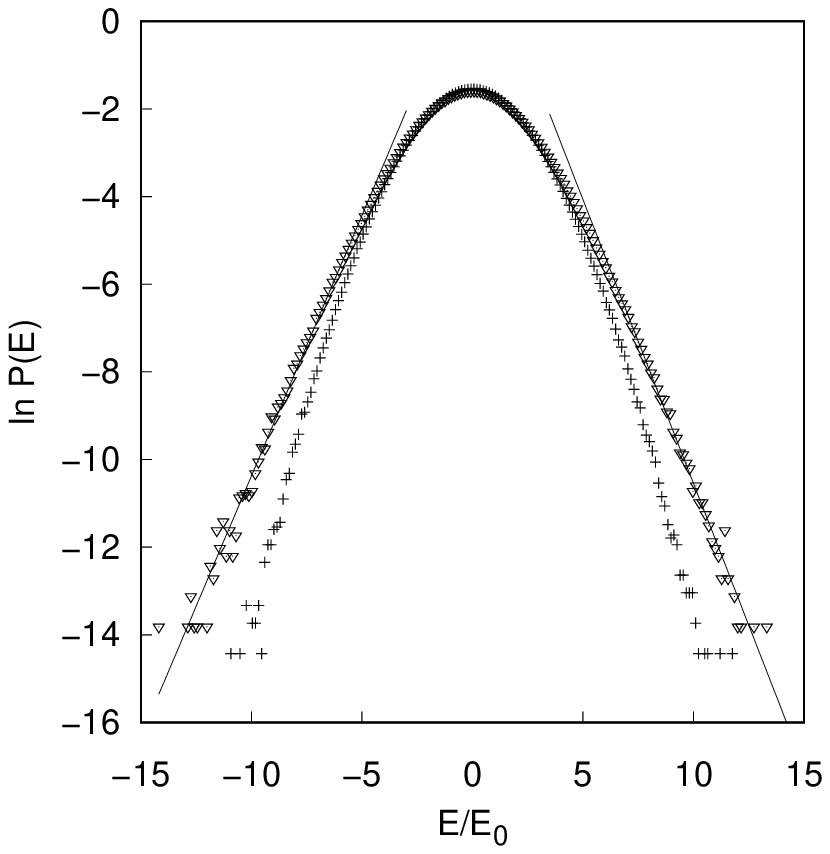}
%\vskip10pt

b)
\caption{Emergence of the exponential tails of the DOS in the axial (a) and planar (b) QG for $l_a=2.9$ ($+$) and $l_a=10$  ($\triangledown$), $L=50$. Central region of the DOS is approximately Gaussian with the same width irrespectively of $l_a$. Straight lines show the best exponential fits for the tails (plotted only for a well developed exponential asymptotics).}
\label{fig-AP}
\end{figure}

\section{Results and Discussion}

\subsection{Central peak of the DOS}

For the absolutely random model of the QG it was already found that the DOS is mostly Gaussian, though for the axial QG the low energy side shows deviation from the Gaussian shape for rather moderate $E$ (decays faster than the Gaussian one), while for the planar QG the DOS is symmetric and retains the Gaussian shape in a more broad region.\cite{Novikov:164510} We describe here the behavior of the low and high energy sides of the DOS assuming $eQ > 0$, for the opposite sign of the carrier the sides are interchanged.

If we consider the main body of the DOS for the locally ordered material, then the variance should be approximately unaffected by $l_a$. Indeed, for the totally random QG  \cite{Novikov:164510}
\begin{equation}\label{sigma2}
  \sigma \propto \frac{eQ}{\varepsilon a^3}.
\end{equation}
Correlated glass could be approximately considered as the random lattice having scale $a l_a$ and built by quadrupoles with the quadrupole moment $Q l_a^3$, thus $\sigma$ remains approximately constant. This consideration should be valid for for moderate and large $l_a$ and is indeed supported by the simulation data (Fig. \ref{s2fig}, inset).

For the random glass the variance $\sigma^2$ is proportional to the only quadratic invariant $\sum\limits_{i,j} Q^2_{ij}$, so for the same value of $Q$ there is a simple relation between variance for axial and planar QG
\begin{equation}\label{s2rel}
  \sigma^2_a = \frac{3}{4}\sigma^2_p,
\end{equation}
and the previous consideration hints that it should be valid for locally ordered glasses, too. Inset of Fig. \ref{s2fig} shows that this is indeed so. Universality means also that the main body of the distribution should have approximately the same shape, and it indeed retains the Gaussian shape (Fig. \ref{s2fig}).

\subsection{Emergence of exponential tails}

\begin{figure}[tbp]
\includegraphics[width=3.375in]{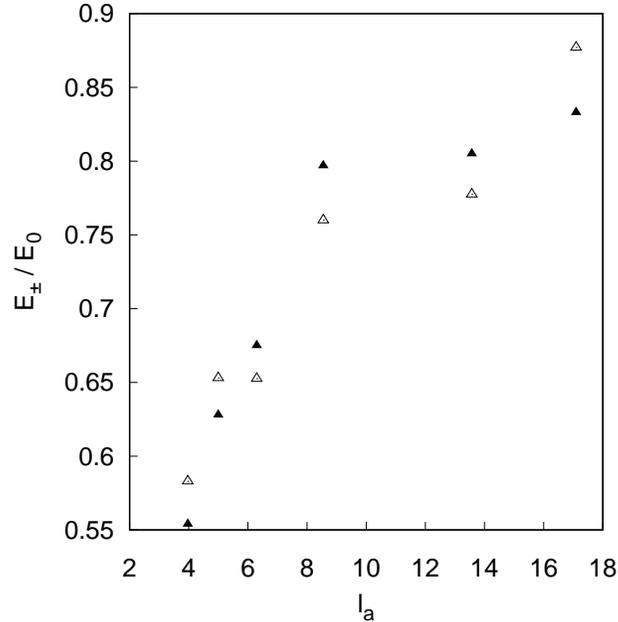}
\caption{Characteristic decay parameters (equivalents of $E_e$ from \eq{exp} for the corresponding tails) of the exponential tails (for $E\rightarrow \pm \infty$) for planar ($\vartriangle$, here $E_+ \approx E_-$) and axial ($\blacktriangle$, only $E_+$ is shown) QG. Low energy tail for the axial glass is approximately Gaussian for rather high $l_a$. Values for $l_a\simeq 3-4$ are not very reliable due to poor statistics.}
\label{Upm}
\end{figure}

The most prominent feature of the DOS in locally ordered QG is the development of the exponential tails for moderate degree of order (see Fig \ref{fig-AP}). Axial QGs demonstrate the high energy tail which is rather easily transforming to the exponential one and the low energy tail which is much more stable and retains its Gaussian shape even for rather large $l_a$ (Fig. \ref{fig-AP}a). Characteristic decay parameter of the tail grows with the increase of $l_a$ (Fig. \ref{Upm}). For low $l_a$ (e.g, for $l_a \le 3$) there is no indication of the development of the exponential tails in the reachable region of $E$.

%fig. 4

\begin{figure}[tbp]
\includegraphics[width=3.375in]{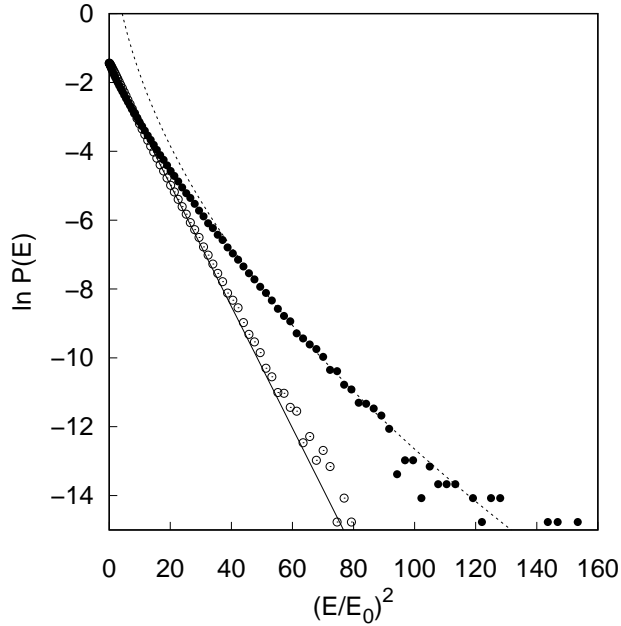}
\caption{DOS for the axial QG with $l_a=5$: low($\circ$) and high ($\bullet$) energy tails, correspondingly. Solid line shows the Gaussian fit for the central peak and the broken line shows the exponential fit.}
\label{U2plot}
\end{figure}

Probably, more convincing evidence for the gradual formation of the exponential tails is the DOS plot as $\ln P(E)$ vs $E^2$ (Fig. \ref{U2plot}, axial QG). Low $E$ tail is almost perfectly Gaussian as a result of the competition of two factors acting in the opposite directions: initially, in the random axial QG the low $E$ tail decays faster than the Gaussian one, but the local order leads to a more slow decay, so the resulting tail becomes more close to the true Gaussian shape. High $E$ tail demonstrates a clear development of the exponential tail.

%fig. 5

\subsection{Spatial distribution of the deep states}

\begin{figure}[tbp]
\includegraphics[width=3.375in]{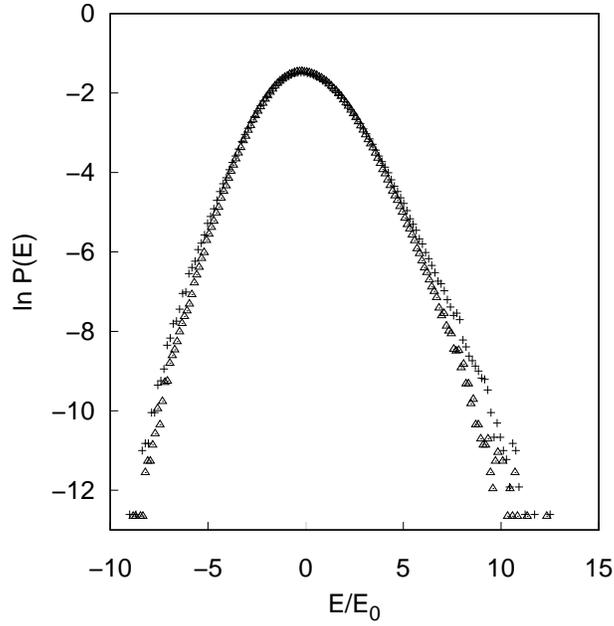}
\caption{Comparison of the DOS for the axial QG and $l_a=7$ for the Voronoi ($\vartriangle$) and small cubes ($+$) models.}
\label{small-cube}
\end{figure}

\begin{figure}[tbp]
\includegraphics[width=3.375in]{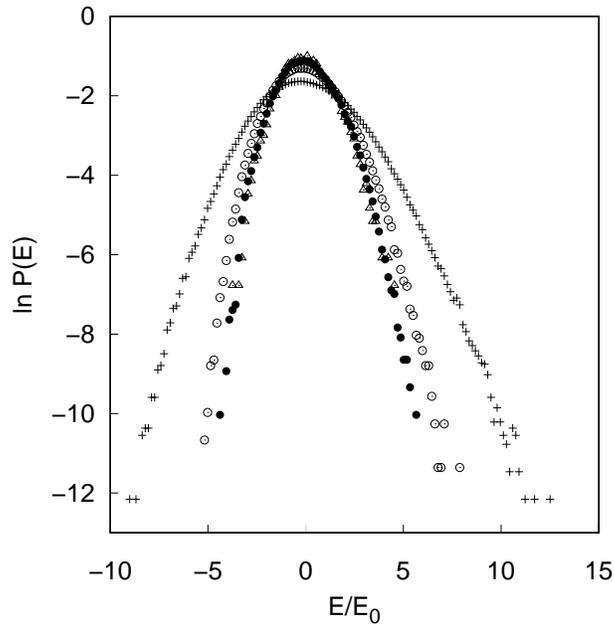}
\caption{Partial DOS for the small cubes model of the axial QG with $l_c=7$ for states with various distance to the cube surface: 0 ($+$), 1 ($\circ$), 2 ($\bullet$), and 3 ($\vartriangle$) lattice scales, correspondingly.}
\label{shells}
\end{figure}

It turns out that the more simple model of the local order where the microstructure is provided by the small cubic chunks of quadrupoles with the linear size of $l_a$ having the same orientation gives approximstely the same DOS (see, for example, the comparison in Fig. \ref{small-cube}). This analogy gives us a possibility to carry out a simple analysis of the spacial location of the deep or, equivalently, very high states. Such states are mostly located in the outer regions of the small cubes (Fig. \ref{shells}).

\begin{figure}[tbp]
\includegraphics[width=3.375in]{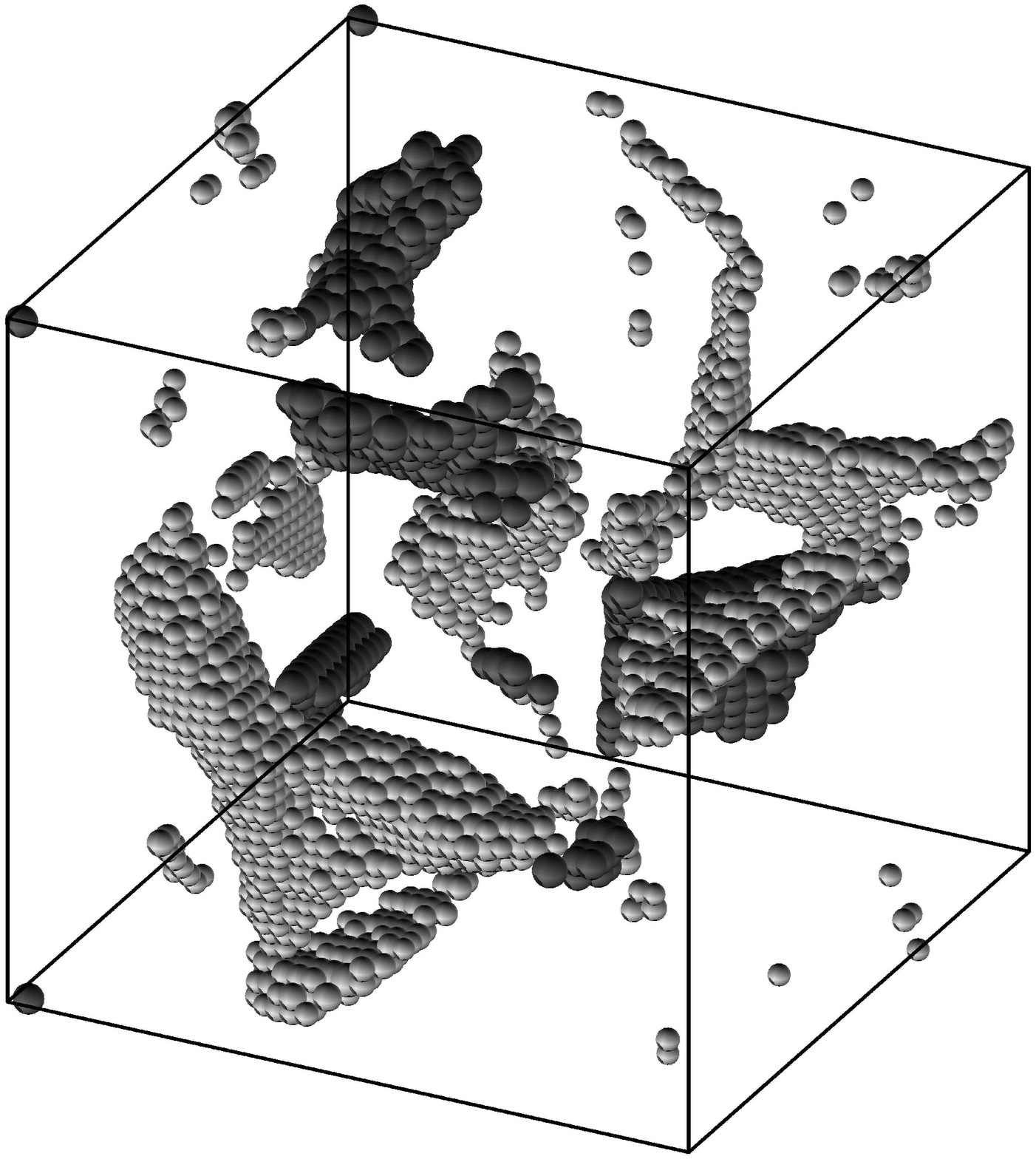}
\caption{Distribution of the deepest (light color) and highest (dark color) states ($|E| \ge 0.5 E_{max}$, $E_{max}$ is the maximal energy for this particular sample) in the $L=50$ cubic sample of the axial Voronoi QG for $l_a=17$, high value of $l_a$ was chosen for the better visibility of the structure. Radius of the ball is proportional to $|E|$. Concentration of the deepest/highest states at the interfaces (outer shells of the Voronoi polyhedra) is clearly visible.}
\label{channels}
\end{figure}

This analysis shows that the spatial distribution of deep states is very different in the locally ordered organic semiconductors in comparison to the totally disordered materials. In disordered semiconductors with electrostatic DOS (dipolar or quadrupolar) there is a cluster structure in the distribution of random energies, and in the cluster the absolute value of $E$ increases from the outer region into the inner part of the cluster. \cite{Novikov:14573,Novikov:41139} In  semiconductors with the structural local order the deepest states are located at the interfaces between ordered granules, and these states form clusters of sites with close energies. Clusters are spreading along the interfaces (Fig. \ref{channels}). Thus, charge transport occurs over 2D random manifold of interfaces with occasional incursions into the bulk of the granules.

%fig. 6

%fig. 7

%fig. 8

\begin{figure}[tbp]
\includegraphics[width=3.375in]{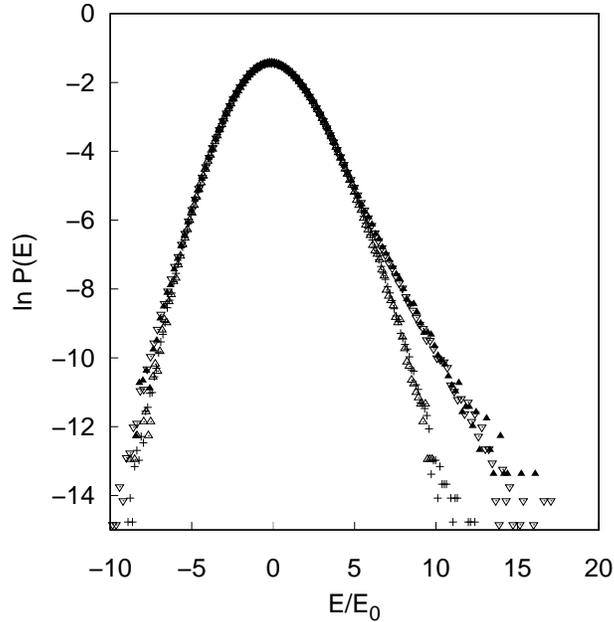}
\caption{DOS for the axial Voronoi QG ($l_a=5$) for $L=50$, $m=0$ ($+$), $L=50$, $m=3$ ($\triangledown$), $L=100$, $m=0$ ($\vartriangle$), and $L=100$, $m=3$ ($\blacktriangle$), correspondingly.}
\label{smooth}
\end{figure}

To test the importance of the sharp interfaces between Voronoi cells for the emergence of the exponential tails we performed a limited test of the DOS features for the smoothed model of the interface (Fig. \ref{smooth}). Spatial orientations of the quadrupole axes for the sharp and smooth model of the interface is shown in Fig. \ref{axes}. We see that the sharp interface is not a necessary requirement of the formation of the exponential tail; in fact, for the smooth interface the tails become even more pronounced. Additional analysis shows that for the smooth version of the small cube model the deepest states again are located at the interfaces. We may conclude that the necessary condition for the exponential tails formation is the parallel orientation of quadrupoles at short distance. Fig. \ref{smooth} provides an additional support for the sufficiency of $L=50$ for the proper evaluation of the DOS.

%fig. 9

\subsection{Locally ordered dipole glasses}

\begin{figure}[tbp]
\includegraphics[width=3.375in]{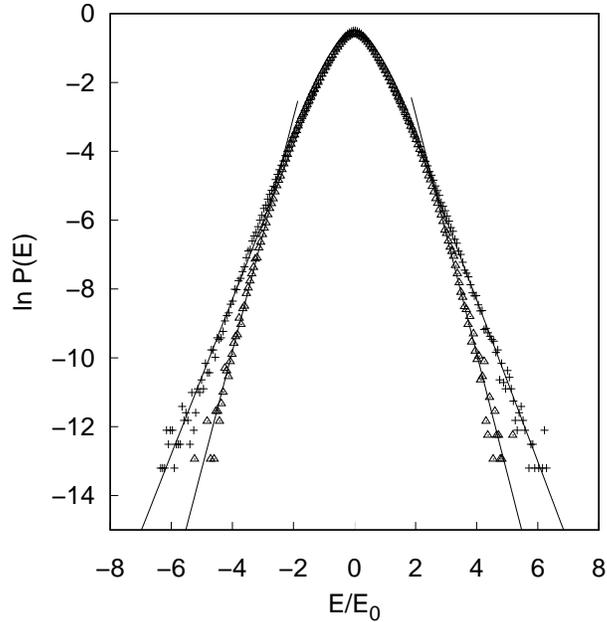}
\caption{DOS for the Voronoi dipole glass for $l_a=5$ ($\vartriangle$) and $l_a=8.5$ ($+$, in both cases $L=50$). Straight lines show the best exponential fit for the tails.}
\label{dip}
\end{figure}

Until now we considered locally ordered quadrupolar glasses, and it turned out that the local order for moderate or large $l_a$ leads to the development of the exponential tails of the DOS. Each quadrupole could be considered as a pair of dipoles with equal and oppositely  oriented dipole moments. We may expect that the dipole glass with anti-ferroelectric local order could mimic the QG and develop exponential tails as well. One of the simplest models demonstrating such behavior could be built in two steps. First, we set up the Voronoi model with dipoles having the same orientation in each cell but orientations in different cells are totally uncorrelated and random (the exact analog of the Voronoi QG). Next, we iteratively flip orientation of each dipole $\vec{p} \rightleftarrows -\vec{p}$ in order to minimize the scalar product $\vec{p}\cdot \sum \vec{p}_i$, where $\vec{p}_i$ are dipoles in the nearest neighborhood of $\vec{p}$. We stop  flipping when the stationary state is reached. Fig. \ref{dip} shows the DOS for the resulting dipole glass.

%fig. 10

As in the case of QG glass, the central peak weakly depends on $l_a$, but the rms disorder is strongly reduced in comparison to the case of totally random dipole glass, where $\sigma\approx 2.35 ep/\varepsilon a^2$, while for Fig. \ref{dip} $\sigma\approx 0.8 ep/\varepsilon a^2$. This is a good illustration of the possible reduction of $\sigma$ due to local ordering. The DOS is symmetric in close analogy with the planar QG.

We are not going to study the DOS in the dipole glasses in details, just to demonstrate that the development of the exponential tails is possible, albeit for the more strict limitations on the local structure of the glass (anti-ferroelectric correlations).

\subsection{Transition region between the Gaussian peak and exponential tail}

A principal difference between the Gaussian and exponential DOS is the energence of the dispersive transport regime for the latter at low temperature $kT < E_e$, where $E_e$ is the characteristic decay parameter for the tail. At the same time, if the exponential tail develops for too low energy, then for the moderately thick transport layer the true dispersive regime may be not observed due to the rather short transit time.

Let $E_t$ be the estimation for the transition region from the Gaussian dependence to the exponential one. Assuming the smooth transition between the Gaussian and exponential regions, $E_t$ may be estimated as
\begin{equation}\label{Ut}
\frac{E_t}{\sigma^2}\simeq \frac{1}{E_e}.
\end{equation}
For time-of-flight experiments in the quasi-equilibrium regime the exponential tail becomes relevant if
\begin{equation}\label{trans}
|E_{eq}|=\frac{\sigma^2}{kT}\simeq E_t\simeq \frac{\sigma^2}{E_e}.
\end{equation}
This relation indicates that the exponential tail becomes relevant if $kT \simeq E_e$, i.e. in the vicinity of the transition to the dispersive transport regime. Observation of the nondispersive transport regime which is governed by the exponential DOS should be difficult. This conclusion should be considered with great care because all estimates here are very approximate.

Assuming typical $\sigma\simeq 0.1$ eV, at the room temperature $kT_r\approx 0.025$ eV $\simeq \sigma/4$ where is no indication of the dispersive transport (hence, $kT_r > E_e$) if
\begin{equation}\label{no-disp}
E_e\simeq \frac{\sigma^2}{E_t} < kT_r, \hskip10pt E_t \gtrsim 4\sigma.
\end{equation}
For a wide range of $l_a$ the rms disorder can be estimated as $\sigma \simeq 1.5-2 E_0$, so $E_t \gtrsim 6-8 \hskip2pt E_0$. Comparison with the DOS figures suggests that for the room temperature the absence of the dispersive regime implies $l_a \le 3-4$.

This brief consideration shows that the description and analysis of experimental time-of-flight curves is, probably, even more difficult and ambiguous than suggested previously; \cite{Schein:7295,Dunlap:9076} emergence of the exponential tails makes transport behavior much more complicated.

An important open problem is the continuation of the exponential tail to an arbitrary deep energy or a possible emergence of another asymptotics for very deep states. Computer simulation cannot solve this problem due to enormous simulation time necessary for the accumulation of the reliable tail statistics. Unfortunately, analytic approach to the estimation of the tail asymptotics is not developed for locally ordered organic glasses.

\section{Conclusions}

In conclusion, we consider the simple model of quadrupolar glasses having local order and show that for the linear spatial size of the ordered regions $\gtrsim 5$ molecules well defined exponential tails are developed. There is a principal difference between axial and planar quadrupoles: for axial QG the DOS is asymmetric and the low energy tail is less prone to the development of the exponential asymptotics, while for the planar QG the DOS is symmetric. For this reason charge transport of electrons and holes in the axial QG may be very different due to principal difference of the shape of the tails. Deep sites, where carriers spend the most time, are organized in clusters located at the interfaces between ordered granules; hence, in most cases charge carrier transport occurs over the random 2D manifold of interfaces.

Comparison of various models suggests that the details of the ordering are not important: what is important is the existence of the domains with preferable parallel orientation of quadrupoles. For dipolar materials the tail is developed for the anti-ferroelectric local ordering. Primary potential candidates for the amorphous organic semiconductors with significant local order are liquid crystalline materials \cite{Garnier:3334,McCulloch:328,Bushby:2012,Eichhorn:88,Woon:2311}  or macrocyclic aromatic hydrocarbons, \cite{Nakanishi:5435} but other materials could develop the sizable local order as well, such as molecularly doped polymers \cite{Parashchuk:1298} or low molecular weight organic glasses. \cite{Thurzo:1108} Quite probably, the dispersive charge transport observed in some organic materials \cite{Dunlap:9076} may be related to the development of the exponential tails due to effects of the local order.

\section*{Acknowledgements}

Financial support from the Ministry of Science and Higher Education of the Russian Federation (A.N. Frumkin Institute) and Program of Basic Research of the National Research University Higher School of Economics is gratefully acknowledged.

\section*{Data availability}

The data that support the findings of this study are available from the corresponding author upon reasonable request.

%\bibliography{novikov}
%

\end{document}